\documentclass{ws-procs975x65}
\usepackage{graphicx}
\usepackage{latexsym}
\usepackage{amssymb}
\def\be{\begin{equation}}
\def\ee{\end{equation}}
\def\bea{\begin{eqnarray}}
\def\eea{\end{eqnarray}}

\begin{document}
\title{Quasi-homogeneous black hole geometrothermodynamics in Einstein-Maxwell theory}

\author{Hernando {Quevedo}}
%\email[]{quevedo@nucleares.unam.mx}
\address{Instituto de Ciencias Nucleares, Universidad Nacional Aut\`onoma de M\`exico, Mexico }
\address{Dipartimento di Fisica and ICRA, Universit\`a di Roma “La Sapienza”, Roma, Italy}
\address{Al-Farabi Kazakh National University, Al-Farabi Ave., 71, Almaty, 050040, Kazakhstan}

\begin{abstract}
In this review, we establish the mathematical framework of geometrothermodynamics (GTD) as a  formalism capable of describing non-extensive, quasi-homogeneous, self-gravitating systems in a Legendre-invariant manner. We argue that the fundamental equations of black holes are quasi-homogeneous functions, a property that invalidates the standard Euler identity of laboratory thermodynamics. We derive the  metrics for the equilibrium manifold and analyze their curvature singularities for the Reissner-Nordström, Kerr, and Kerr-Newman black holes. Furthermore, we establish a direct  correspondence between the curvature singularities of the equilibrium space and phase transitions, as determined by the divergences of the corresponding heat capacities.
\end{abstract}

\section{Introduction}

The thermodynamics of black holes has remained one of the most fertile and challenging fields of theoretical physics since the pioneering works of Bekenstein and Hawking in the 1970s. The realization that black holes possess a temperature $T$ and an entropy $S$ led to the formulation of the four laws of black hole mechanics, which bear a striking mathematical resemblance to the laws of classical thermodynamics. However, a profound conceptual crisis arises when one attempts to treat a black hole as a standard thermodynamic system. Unlike laboratory systems, where forces are short-ranged and variables are additive, gravity is a long-range interaction that leads to the inherent non-extensivity of the system.

In classical thermodynamics, a fundamental equation $\Phi = \Phi(E^1, \dots, E^n)$ is expected to be a homogeneous function of degree one. This homogeneity ensures the validity of the Euler identity and, consequently, the Gibbs-Duhem relation, which allows the intensive variables to be related to one another. For a black hole, however, the entropy scales with the area of the event horizon ($S \propto M^2$ in the Schwarzschild case), implying that the mass is not a linear function of its extensive parameters. If one persists in applying the standard Euler identity to a Schwarzschild or Kerr black hole, the result is a vanishing Gibbs-Duhem relation, which effectively removes the possibility of defining a consistent phase space for phase transitions.

To resolve this inconsistency, we adopt the framework of {quasi-homogeneous thermodynamics}, a concept originally discussed by Belgiorno \cite{Belgiorno2003} and rigorously formalized in reference  \cite{Quevedo2019}. In this paradigm, the thermodynamic potential is treated as a quasi-homogeneous function, where each extensive variable is assigned a specific weight $\beta_a$. This mathematical shift allows for a generalized Euler identity that accounts for the scaling properties of the gravitational field, thereby restoring the degrees of freedom necessary for a robust thermodynamic description.

Furthermore, any geometric representation of these states must satisfy the principle of {Legendre invariance}. Classical geometric approaches, such as those proposed by Weinhold \cite{wei1} or Ruppeiner \cite{rup79,rup95}, utilize the Hessian of the internal energy or entropy to define a metric. While these metrics provide insights into the fluctuations of the system, they are not invariant under a change of thermodynamic potential. A physical theory of black hole equilibrium should not depend on whether one chooses the mass (enthalpy), the temperature (internal energy), or the Gibbs free energy as the starting potential.

{Geometrothermodynamics (GTD)} emerged as a response to these requirements. Using the differential geometry of contact manifolds, GTD constructs a phase space where Legendre transformations are treated as coordinate changes that preserve the underlying contact structure. On the equilibrium manifold, GTD introduces metrics that explicitly incorporate the quasi-homogeneity of the potential. This ensures that the resulting thermodynamic curvature scalar $R$ is a true invariant, whose singularities act as unambiguous indicators of phase transitions. This paper aims to detail this connection, proving that the quasi-homogeneous nature of black holes is not merely a mathematical curiosity but a physical property that is necessary to build a consistent geometrothermodynamic formalism.

\section{Black hole thermodynamics}

Under the no-hair theorems of Einstein-Maxwell theory, electro-vacuum black holes are uniquely determined by a triplet of parameters: mass $M$, angular momentum $J$, and electric charge $Q$. The geometry of the associated gravitational field is provided by the Kerr-Newman metric, which, utilizing Boyer-Lindquist coordinates, takes the form \cite{solutions}
\bea
ds^2 = &-&\frac{\Delta - a^2\sin^2\theta}{\Sigma} dt^2 
-\frac{2a\sin^2\theta (r^2+a^2 -\Delta)}{\Sigma} dtd \varphi \nonumber\\
& +& \frac{(r^2+a^2)^2 - a^2\sin^2\theta\, \Delta }{\Sigma} \sin^2\theta d\varphi^2
+\frac{\Sigma}{\Delta} dr^2 + \Sigma d\theta^2 \ , 
\label{kn}
\eea
\be 
\Sigma = r^2 + a^2\cos^2\theta\ , \quad \Delta = (r-r_+)(r-r_-)\ ,
 \quad r_\pm = M \pm \sqrt{M^2-a^2-Q^2} \ ,
 \label{horkn}
\ee
where $ a = J/M$ is the specific angular momentum. 
In 1973, Bekenstein \cite{bek73} established that the horizon area $A$ of a black hole corresponds to the entropy $S$ of a classical thermodynamic system. This discovery marked the inception of black hole thermodynamics \cite{bch73,haw75,davies}. While its underlying statistical mechanics remains elusive, this field has seen extensive investigation over the last thirty years, driven largely by its potential link to a prospective theory of quantum gravity.

In the framework of black hole thermodynamics, it has been proved that the physical parameters that enter the Kerr-Newman black hole 
satisfy the first law of  
thermodynamics \cite{bch73}
\be 
dM = T dS +  \phi d Q +\Omega_H  d J  \ ,
\label{first}
\ee
where  $T$ is the Hawking temperature \cite{haw75} which is proportional to the surface 
gravity on the horizon, $S=A/4$ is the entropy,
 $\Omega_H $ is the angular velocity on the horizon, 
and $\phi$ is the electric potential. As in ordinary thermodynamics, all the 
thermodynamic information is contained in the fundamental equation which was 
first derived by Smarr \cite{smarr73} 
\be
M = \left[ \frac{\pi J^2}{S} + \frac{S}{4\pi}\left(1 + \frac{\pi Q^2}{S}\right)^2\right]^{1/2}\ .
\label{feqbh}
\ee 
In the entropy representation, this fundamental equation can be rewritten as
\be
S= \pi\left( 2 M^2 - Q^2 +2\sqrt{M^4-M^2Q^2 - J^2}\right) \ .
\label{entropy}
\ee 
Davies \cite{davies} argued 
that black holes undergo a second order phase transition at the points where the heat
capacity diverges. This argument is supported by the result that some critical exponents
related to the singular points obey scaling laws \cite{sokmaz,lau1,lau2,munpir,lou1,lou2}.
Following Davies, we assume in this work that the structure of the phase transitions 
of the Kerr-Newman black hole is determined by the corresponding heat capacity 
\be
C_{Q,J}=T\left(\frac{\partial S}{\partial T}\right)_{Q,J} = \left(\frac{\frac{\partial M}{\partial S}}{\frac{\partial^2 M}{\partial M^2}}\right)_{Q,J}. 
\ee
Then,
\be
C_{Q,J} = -\frac{4TM^3S^3}
{2M^6-3M^4Q^2-6M^2J^2+Q^2J^2 + 2(M^4-M^2Q^2 - J^2)^{3/2} } \ ,
\label{capkn}
\ee
where the Hawking temperature is giveven by
\be
T={\frac {\sqrt {{M}^{4}-{Q}^{2}{M}^{2}-{J}^{2}}}{2\pi \,M \left( 2
\,{M}^{2}-{Q}^{2}+2\,\sqrt {{M}^{4}-{Q}^{2}{M}^{2}-{J}^{2}} \right) }}
.
\ee
Consequently, the phase transition structure of the Kerr-Newman black hole is determined by the solutions of the equation 
\be
{2M^6-3M^4Q^2-6M^2J^2+Q^2J^2 + 2(M^4-M^2Q^2 - J^2)^{3/2} } =0.
\ee
In the following sections, we will see that equivalent conditions can be obtained by using the formalism of GTD.

\section{Review of geometrothermodynamics}
\label{sec:gtd}

The geometrothermodynamics (GTD) framework is designed to represent the characteristics of thermodynamic systems through differential geometry \cite{quev07}. Within this approach, each system is associated with an equilibrium space ${\cal E}$, where individual points correspond to specific equilibrium states. This space is defined as a differentiable structure, allowing it to be endowed with a Riemannian metric $g$. Consequently, the Riemannian manifold $({\cal E}, g)$ provides a geometric description that reflects the underlying thermodynamics; specifically, geodesics within ${\cal E}$ represent quasistatic processes, and divergences in the curvature scalar are identified with the occurrence of phase transitions \cite{aqs08}.

A central feature of GTD is its invariance under Legendre transformations, a symmetry that reflects the thermodynamic principle that a system's description should not depend on the specific choice of potential. In classical thermodynamics, these potentials are interconnected via Legendre maps \cite{callen}. GTD incorporates this property by defining an auxiliary $(2n+1)$-dimensional phase space ${\cal T}$, in which Legendre transformations are represented as diffeomorphisms. This phase space is equipped with a Riemannian metric $G$ that is strictly required to maintain invariance under such transformations.

The construction of $G$ is based on elements of the contact geometry. As a general result, three classes of Legendre-invariant metrics arise, which we denote as $G^I$, $G^{II}$, and $G^{III}$. Defining the coordinates of ${\cal T}$ by $Z^A = \{\Phi, E^a, I_a\}$ with $a = 1, \ldots, n$, the GTD metrics of the phase space can be written as
\begin{equation}
    G^{I/II} = \left(d\Phi - I_a \, dE^a \right)^2 + (\xi_{ab} E^a I^b)(\chi_{cd} \, dE^c dI^d),
\end{equation}
\begin{equation}
    G^{III} = \left(d\Phi - I_a \, dE^a \right)^2 + \sum_{a=1}^{n} \xi_a (E_a I_a)^{2k+1} \, dE^a dI^a.
\end{equation}
Here, $\xi_a$ are real constants, $\xi_{ab}$ is a diagonal $n \times n$ real matrix, and $k$ is an integer. The matrix $\chi_{cd}$ is defined as $\delta_{cd} = \mathrm{diag}(1,1,\ldots,1)$ for $G^I$, and $\eta_{cd} = \mathrm{diag}(-1,1,\ldots,1)$ for $G^{II}$. The matrix $\delta_{ab}$ is used to raise and lower indices of $E^a$ and $I_a$.

It is crucial to note that the explicit structure of the GTD metrics is derived by imposing that the components of $G = G_{AB} \, dZ^A dZ^B$ are unaffected by Legendre transformations. From a physical standpoint, this requirement guarantees that the thermodynamic behavior of a system is independent of the specific potential chosen for its representation. By integrating this essential property of classical thermodynamics \cite{callen} directly into the GTD formalism, a more comprehensive description of phase transition structures is achieved, particularly in the realm of black hole thermodynamics.

Correspondingly, the equilibrium space $({\cal E}, g)$ is defined as an $n$-dimensional subspace of the phase space $({\cal T}, G)$, which inherits the core property of Legendre invariance. More precisely, if $E^a$ represents the coordinates of the equilibrium space, then $\Phi(E^a)$ denotes any thermodynamic potential accessible via Legendre transformations within the energy or entropy representations. Consequently, the set $E^a$ may consist of any combination of $n$ independent thermodynamic variables and their conjugate partners. This ensures that the outcomes of the GTD framework remain entirely independent of the chosen thermodynamic potential and its associated independent variables.

Furthermore, in GTD, any thermodynamic potential can be considered as the fundamental equation, $\Phi = \Phi(E^a)$, which describes the system and satisfies the first law
\begin{equation}
d\Phi = I_a \, dE^a , \qquad
I_a = \frac{\partial \Phi}{\partial E^a} .
\end{equation}

As previously noted, quasi-homogeneity is a vital requirement for achieving consistent results in black hole thermodynamics. We therefore impose this same condition on the potential $\Phi(E^a)$, assuming that
\be
\Phi(\lambda^{\beta_a} E^a) = \lambda^{\beta_\Phi} \Phi (E^a) .
\ee
This property is essential in the GTD framework, as the quasi-homogeneity coefficients are directly incorporated into the explicit structure of the equilibrium space metrics. Furthermore, it turns out that to ensure that the three metrics of the equilibrium space can be applied consistently to the same system, we must set $k=0$ in the phase space metric  $G^{III}$.

The above consistency conditions allow us to express the GTD metrics of the equilibrium space as
\begin{equation}
g^{I} = \sum_{a,b,c=1}^{n} \left( \beta_c E^c \frac{\partial \Phi}{\partial E^c} \right)
\frac{\partial^2 \Phi}{\partial E^a \partial E^b} \, dE^a dE^b ,
\label{gdownI}
\end{equation}
\begin{equation}
g^{II} = \sum_{a,b,c,d=1}^{n} \left( \beta_c E^c \frac{\partial \Phi}{\partial E^c} \right)
\eta_{a}^{\ d}
\frac{\partial^2 \Phi}{\partial E^b \partial E^d} \, dE^a dE^b ,
\label{gdownII}
\end{equation}
\begin{equation}
g^{III} = \sum_{a,b=1}^{n} \left( \beta_a E^a \frac{\partial \Phi}{\partial E^a} \right)
\frac{\partial^2 \Phi}{\partial E^a \partial E^b}
\, dE^a dE^b ,
\label{gdownIII}
\end{equation}
where $\eta_a^{\ c} = \mathrm{diag}(-1,1,\ldots,1)$.

Moreover, to simplify calculations, we can utilize the quasi-homogeneous Euler identity and the Gibbs--Duhem relation \cite{Quevedo2019}, which are expressed as
\be
\sum_{a=1}^n \beta_a I_a E^a = \beta_\Phi \Phi ,
\quad
\sum_{a=1}^n \left[ (\beta_a-\beta_\Phi) I_a dE^a + \beta_a E^a dI_a\right] = 0.
\label{identities}
\ee
These expressions reduce to the standard homogeneous identities when $\beta_a = \beta_\Phi = 1$.

It is crucial to note that the explicit forms of the GTD metrics, defined in Eqs.~(\ref{gdownI})--(\ref{gdownIII}), were not developed to mimic the behavior of any specific thermodynamic model. Rather, their derivation stems from the objective of geometrically embedding the physical requirement that a system's thermodynamic characteristics remain invariant under the choice of potential. The consistency of this purely geometric formulation with other physical laws, as well as its ability to offer novel insights, is an essential question. We shall demonstrate that the GTD framework significantly enhances our comprehension of black hole thermodynamics.

To this end, consider the case $n=2$ with fundamental equation $\Phi = \Phi(E^1, E^2)$. The corresponding line elements can be written as 
\begin{align}
g^{I} &= \Sigma \left[\Phi_{,11} (dE^1)^2 + 2 \Phi_{,12} dE^1 dE^2 + \Phi_{,22} (dE^2)^2 \right], 
\label{gI2D} \\
g^{II} &= \Sigma \left[-\Phi_{,11} (dE^1)^2 + \Phi_{,22} (dE^2)^2 \right],
\label{gII2D} \\
g^{III} &= \beta_1 E^1 \Phi_{,1} \Phi_{,11} (dE^1)^2 
+ \Sigma \Phi_{,12} dE^1 dE^2
+ \beta_2 E^2 \Phi_{,2} \Phi_{,22} (dE^2)^2 ,
\label{gIII2D}
\end{align}
where $\Phi_{,a} = \partial \Phi / \partial E^a$ and 
\[
\Sigma = \beta_1 E^1 \Phi_{,1} + \beta_2 E^2 \Phi_{,2}.
\]

Using the Euler identity, $\Sigma = \beta_\Phi \Phi$, the curvature scalars indicate the presence of  singularities, which are determined by the conditions
\begin{equation}
\text{I:} \quad \Phi_{,11}\Phi_{,22} - (\Phi_{,12})^2 = 0 ,
\label{condI}
\end{equation}
\begin{equation}
\text{II:} \quad \Phi_{,11} \Phi_{,22} = 0 ,
\label{condII}
\end{equation}
\begin{equation}
\text{III:} \quad \Phi_{,12} = 0 .
\label{condIII}
\end{equation}
Remarkably, these conditions coincide with the stability conditions  of a system with two thermodynamic degrees of freedom \cite{callen}  and, consequently, can determine the phase transition structure of the system. This result explicitly shows that in GTD the curvature singularities are directly related to  phase transitions. 

In the original formulation of black hole thermodynamics \cite{davies}, it is assumed that divergences of the heat capacity are associated with second-order phase transitions. In practice, the divergences of the heat capacity, determined by the zeros of the second derivative of the thermodynamic potential, e.g., $\Phi_{,11}=0$ for $n=2$. This condition is reproduced in GTD by  condition~(\ref{condII}). However, as follows from Eqs.~(\ref{condI})--(\ref{condIII}), GTD predicts additional types of phase transitions associated with $\Phi_{,22}=0$, $\Phi_{,12}=0$, and condition~(\ref{condI}), which correspond to divergences of other response functions and to violations of thermodynamic stability \cite{callen}. Similar results hold for systems with $n \geq 3$. 

This highlights that the GTD framework identifies a more intricate phase transition landscape in black hole thermodynamics, significantly broadening the scope of the classical criteria established by Davies \cite{davies}.

Quasi-homogeneity represents another significant contribution of GTD to the study of black hole thermodynamics. While this property had been explored in alternative contexts \cite{Belgiorno2003}, GTD identifies it as a mandatory requirement for the consistent application of its three metrics to a single system and as a cornerstone of the thermodynamic description of black holes. Consequently, gravitational coupling constants—such as the cosmological constant or the Gauss--Bonnet parameter—must be interpreted as independent thermodynamic variables. This shift necessitates an extended thermodynamic framework, offering novel perspectives on the physics of black holes.

Lastly, recent data from black hole shadows and light rings have catalyzed the integration of GTD into the study of shadows, phase transitions, and black hole microstructure . This synergistic approach has paved the way for an emerging field of inquiry known as {shadow geometrothermodynamics}.

\section{Geometrothermodynamics of black holes}
\label{sec:gtdbh}

From the explicit expressions of the metrics of the equilibrium space, it follows that we can derive all the geometric properties from the fundamental equation. In the case of the Kerr-Newman black hole, the entropy (\ref{entropy}) represents the fundamental equation. It is easy to show that this entropy is a quasi-homogeneous function $S(\lambda^{\beta_a} E^a) =\lambda^{\beta_S}S(E^a)$ with $E^a=(M,Q,J)$, if the following relationships are satisfied
\be
\beta_Q=\beta_M, \quad \beta_J = 2\beta_M, \quad \beta_S = 2\beta_M ,
\ee
which prove that in fact the Kerr-Newman black hole is a quasi-homogeneous thermodynamic system. Thus, we can apply the results of the previous section. 

First, we will investigate the case of two-dimensional GTD, which correspond to the Reissner-Nordström and Kerr black holes, Then, we will proceed with the Kerr-Newman black hole in three-dimensional GTD. It turns out that the phase transition structure, as dictated by the behavior of the heat capacity, is entirely contained in the metric $g^{II}$. Therefore, we will consider explicitly only this metric for the equilibrium space.

\subsection{The Reissner-Nordstr\"om black hole}
\label{sec:rn}

The Reissner-Nordström metric is derived from Eq.(\ref{kn}) by setting $J=0$. It characterizes a static, spherically symmetric black hole featuring two horizons located at
\be
r_\pm = M\pm\sqrt{M^2-Q^2}\ . 
\ee
We assume that $Q\leq M$ to avoid the presence of  
naked singularities. 

The entire thermodynamic information of this black hole is contained in 
the fundamental equation which, in the entropy representation, becomes
\be
S=\pi \left(M+\sqrt{M^2-Q^2}\right)^2 \ .
\label{srn}
\ee

According to Davies \cite{davies}, the phase transition structure of the Reissner-Nordstr\"om 
black hole can be derived from the heat capacity
\be
C_{Q} = \frac{4TM^3S^3}
{- 2M^6 + 3M^4 Q^2 - 2(M^4-M^2Q^2)^{3/2} } 
=-\frac{2\pi^2 r_+^2 (r_+ - r_-)}{r_+ - 3r_-} \ .
\label{caprn}
\ee
To proceed with our geometric approach to black hole thermodynamics, we need only the
fundamental equation as given in (\ref{srn}) from which we can calculate the 
thermodynamic metric
\bea
g^{II}_{RN}& = & \beta_S (M S_{,M} + Q S_{,Q})\left(S_{,MM} dM^2 - S_{,QQ}dQ^2 \right)\nonumber\\
&=& \frac{8\pi^2r_+^3}{(r_+-r_-)^3}\left[
2r_+(r_+-3r_-) dM^2 - (r_+^2 + 3r_-^2)dQ^2\right]
\eea
This metric is singular in the extremal limit $r_+=r_-$, which could lead to a breakdown of our GTD approach. However, 
the analysis of the corresponding scalar curvature  
\be
R^{RN} = \frac{(r_+^2 - 3r_-r_++6r_-^2)(r_++3r_-)(r_+-r_-)^2}{
\pi^2 r_+^3 (r_+^2+3r_-^2)^2 (r_+-3r_-)^2}
\ee
shows that the space of equilibrium states becomes flat in the extremal black hole limit, indicating that it corresponds to a coordinate singularity.
 Moreover, from the
expression for the scalar curvature we see  that the only singular point corresponds
to the value $r_+ = 3 r_-$ which is exactly the point where a phase transition 
occurs in the heat capacity (\ref{caprn}).

%%%%%%%%%%%%%%%%%%%%%%%%%%%%%%%%%%%%%%%%%%%%%%%%%%%%%%%%%%%%%%%
\subsection{The Kerr black hole}
\label{sec:kerr}

The Kerr metric corresponds to the limit $Q=0$ of the Kerr-Newman metric (\ref{kn}).
It describes the gravitational field of a stationary, axially symmetric, rotating
black hole with two horizons situated at 
\be
r_\pm = M \pm \sqrt{M^2 - J^2/M^2}\ .
\ee
The corresponding thermodynamic 
fundamental equation in the entropy representation can be expressed as
\be
S=2\pi\left(M^2 + \sqrt{M^4-J^2}\right) \ .
\ee
According to Davies approach, second order phase transitions occur at the points where the 
heat capacity 
\be
C_{J} = \frac{4TM^3S^3}
{6M^2 J^2 - 2M^6 - 2(M^4-J^2)^{3/2} } 
=  \frac{2\pi^2 r_+ (r_++r_-)^2(r_+-r_-)}{ r_+^2-6r_+r_--3r_-^2}
\label{capkerr}
\ee
diverges. We assume values of the mass in the range $M^2\geq J$, where the equal sign corresponds to  
the extremal limit of the Kerr black hole in which the two horizons coincide.

The Legendre invariant metric reduces in this case to 
\bea
g^{II}_K  & = & \beta_S (MS_{,M} + J S_{,J})\left( S_{,MM} dM^2 - S_{,JJ}dJ^2\right) \nonumber\\
&=& \frac{16\beta_S\pi^2r_+^2 (r_++r_-)}{(r_+-r_-)^4} \left[  r_+(r_+^2-6r_+r_--3r_-^2) dM^2 
- (r_+ + r_-)dJ^2 \right] .
\label{gkerr}
\eea
This  metric has also a coordinate singularity at the extremal limit $r_+=r_-$. 
The scalar curvature for the thermodynamic metric of the Kerr black hole can be 
expressed as 
\be
R^K = \frac{(3r_+^3+3r_+^2 r_- + 17 r_+r_-^2 + 9 r_-^3)(r_+-r_-)^3}
{2\pi^2 r_+^2(r_++r_-)^4(r_+^2-6r_+r_--3r_-^2)^2} \ .
\ee
The curvature singularities are situated at the roots of the 
polynomial  equation $r_+^2-6r_+r_--3r_-^2=0$. 
 These are exactly the zeros that determine the critical
points where phase transitions take place, according to  the 
heat capacity (\ref{capkerr}),

\section{The general Kerr-Newman black hole}
\label{sec:gtd3}

The Kerr-Newman metric (\ref{kn}) describes the gravitational field of a rotating, charged black hole. The horizons are determined by Eq.(\ref{horkn}). According 
to our results of section \ref{sec:gtdbh}, the space of thermodynamic 
equilibrium states is three-dimensional and the corresponding Legendre invariant
metric can be written as
\be
g^{II}_{KN}=\beta_S (MS_{,M }+ QS{,_Q}+ J S_{,J})\left(
S_{,MM}dM^2 - S_{, QQ} dQ^2  - 2S_{,QJ} dQ dJ 
- S_{JJ}dJ^2 \right) ,
\ee 
where the entropy function is given in Eq.(\ref{entropy}). The explicit expression of this metric cannot be written in a compact form. The corresponding curvature scalar can be expressed as
\be
R^{KN} = \frac{N}{D}\ , \quad
D= 4 (MS{,_M} + QS{,_Q}+ J S{,_J})^3 ( S_{,QJ}^2 - S_{,QQ}S_{JJ} ) ^3 S_{,MM}^2
\ee
so that the curvature singularities are determined by the zeros of the function $D$. 
Furthermore, it can be shown that the only nontrivial zeros of this function are determined by the expression
\be
D\propto \left[2M^6-3M^4Q^2-6M^2J^2+Q^2J^2 + 2(M^4-M^2Q^2 - J^2)^{3/2}\right]^2 \ ,
\ee
which is exactly the denominator
of the heat capacity (\ref{capkn}). This proves that the phase transitions correspond to curvature singularities of thermodynamic metric $g^{II}_{KN}$.

\section{Conclusion}
\label{sec:con}

In this work, we have shown that 
quasi-homogeneous geometrothermodynamics provides a consistent, Legendre-invariant description of black holes in Einstein-Maxwell theory. By incorporating the quasi-homogeneous coefficients of the fundamental equation into the metric structure of the equilibrium space, we obtain general expressions for the Legendre invariant metrics that can be applied to any thermodynamic system, either homogeneous or quasi-homogeneous. 

We have shown that the phase transition structure of the Einstein-Maxwell black holes, as determined by the divergences of the heat capacity, correspond to curvature singularities of the metric $g^{II}$ of the equilibrium space.
As for the remaining metrics $g^I$ and $g^{III}$, it can be shown that their curvature singularities correspond to divergences of response functions, other than the heat capacity. The physical interpretation of these additional divergences in black hole thermodynamics will be treated elsewhere.

\end{document}